\documentclass[%
preprint,
reprint,amsmath,amssymb,aps,
superscriptaddress,
preprintnumbers,
nofootinbib, cases,
 amsmath,amssymb,
 aps,
]{revtex4-2}

\usepackage{ulem,graphicx}
\usepackage{dcolumn}
\usepackage{bm}
\usepackage{hyperref}
\usepackage{cleveref}
\usepackage{booktabs}

\usepackage{physics}
\usepackage[english]{babel}
\usepackage[dvipsnames]{xcolor}
\usepackage{xr}
\usepackage{subfigure}
\usepackage{aas_macros}

\newcommand{\steedit}[1]{\textcolor{black}{#1}}

\newcommand{\sdmedit}[1]{\textcolor{black}{#1}}

\begin{document}

\title{Neutron Star Quantum Death by Small Black Holes}

\author{Pierce Giffin}%
\email{pgiffin@ucsc.edu}
 \affiliation{Department of Physics, University of California, Santa Cruz, CA 95064, USA}
\affiliation{Santa Cruz Institute for Particle Physics, Santa Cruz, CA 95064, USA}

\author{John Lloyd}%
\email{jblloyd@ucsc.edu}
 \affiliation{Department of Physics, University of California, Santa Cruz, CA 95064, USA}

\author{Samuel D. McDermott}%
\email{sammcd00@fnal.gov}
 \affiliation{Fermi National Accelerator Laboratory, Batavia, IL 60510 USA}

\author{Stefano Profumo}%
\email{profumo@ucsc.edu}
\affiliation{Department of Physics, University of California, Santa Cruz, CA 95064, USA}
\affiliation{Santa Cruz Institute for Particle Physics, Santa Cruz, CA 95064, USA}

\date{\today}

\begin{abstract}
Neutron stars can be destroyed by black holes at their center accreting material and eventually swallowing the entire star. Here we note that the accretion model adopted in the literature, based on Bondi accretion or variations thereof, is inadequate for small black holes -- black holes whose Schwarzschild radius is comparable to, or smaller than, the neutron's de Broglie wavelength. In this case, quantum mechanical aspects of the accretion process cannot be neglected, and give rise to a completely different accretion rate. We show that for the case of black holes seeded by the collapse of bosonic dark matter, this is the case for electroweak-scale dark matter particles. In the case of fermionic dark matter, typically the black holes that would form at the center of a neutron star are more massive, unless the dark matter particle mass is very large,  larger than about 10$^{10}$ GeV. We calculate the lifetime of neutron stars harboring a ``small'' black hole, and find that black holes lighter than $\sim 10^{11}$ kg quickly evaporate, leaving no trace. More massive black holes destroy neutron stars via quantum accretion on time-scales much shorter than the age of observed neutron stars. \steedit{We find that the range where seed black holes inside neutron stars are massive enough that they do not quickly evaporate away, but not so massive that a fluid accretion picture is warranted is limited to between $\sim10^{11}$ and $10^{12}$ kg, but our results are key to accurately determine the actual critical black hole mass corresponding to the onset of neutron star destruction}
\end{abstract}

\maketitle
\clearpage

\section{Introduction}
The very existence of long-lived neutron stars (NS) imposes significant constraints on dark matter: as was recognized long ago \cite{Goldman:1989nd, Gould:1989gw}, dark matter can be captured and accumulated in NS (if pair-annihilation is sufficiently slow or absent, see e.g. \cite{Zurek:2013wia}), thermalize, and collapse into a ``small'' black hole that could eventually swallow and destroy the NS \cite{Kouvaris:2007ay, Bertone:2007ae, Kouvaris:2010vv, Kouvaris:2011fi, McDermott:2011jp, Bramante:2013hn, Bell:2013xk}. If dark matter consists of primordial black holes (PBH) \cite{Carr:2009jm}, NS may capture PBHs, potentially leading to the disruption of the NS \cite{Pani:2014rca} -- the capture rate is however too small to set any meaningful constraints on PBH as dark matter \cite{Montero_Camacho_2019}.

Thus far, the treatment of NS material accretion onto a BH at the core of a NS has followed the assumption that accretion proceeds through a spherical Bondi-Hoyle process, possibly including caveats from the NS rotation \cite{Kouvaris:2013kra} or from Pauli blocking \cite{Autzen:2014tza} (see also Ref.~\cite{East:2019dxt,Richards:2021upu} for numerical studies of the full general relativistic problem of black hole evolution, but also assuming Bondi accretion). The Bondi-Hoyle accretion picture presupposes spherically symmetric,
steady state accretion of a non-self-gravitating gas \cite{Bondi:1952ni}, which is treated as a fluid with a polytropic equation of state. Here we critically note that {\em this treatment breaks down when the individual particle quantum size -- its de Broglie wavelength -- exceeds the size of the black hole, i.e. its Schwarzschild radius}. In that case, wavelike effects become important, and the absorption cross section is given by the expression in the classic work by Unruh, Ref.~\cite{Unruh:1976fm}. A key assumption in the Bondi picture -- the absence of outflows -- breaks down when particles effectively scatter off, and are not always absorbed by, the hole.

Let us estimate the range of black hole masses when Unruh's treatment is necessary. For simplicity, we treat the NS as consisting of a neutron population modeled as a degenerate Fermi gas with density $n_n\simeq 0.3 \, {\rm fm}^{-3}$, leading to a Fermi momentum
$$p_F=\hbar\left(3\pi^2 n_n\right)^{1/3}\simeq \sdmedit{0.4}\ {\rm GeV},$$ and a corresponding Fermi velocity $$v_F=p_F/E_F\simeq \sdmedit{0.4}c.$$ The NS temperature $T_{\rm NS}\simeq \mathcal O( 10^8) \, {\rm K} \simeq $ \sdmedit{10 k}eV is much lower than the Fermi energy, \sdmedit{and relativistic corrections due to degeneracy-pressure induced velocity are at the level of $\sim 10$\%. Thus,} we model the energy distribution as $f(E)\propto E^{1/2}$ from the density of state for a \sdmedit{nonrelativistic} 3D free electron gas, and the velocity distribution, correspondingly, as $f_F(v)\propto v^2$, and limited to $v\le v_F$, thus 
$$
f_F(v)=\frac{3v^2}{v_F^3},\ v\le v_F;\quad f_F(v)=0,\ v>v_F.
$$
The average velocity of these Fermi-degenerate neutrons is $\langle v \rangle_F=\int v f_F(v)/\int f_F(v) \simeq \sdmedit{0.3} c$, which we will use as a typical velocity below.

The key assumption of the Unruh treatment of ``quantum'' accretion onto a Schwarzschild black hole \cite{Unruh:1976fm} is that the Schwarzschild radius of the hole $R_{\rm Schw} = 2GM_{\rm BH}$ be smaller than the de Broglie wavelength of the particles being absorbed, and that said particle be described as a free plane wave asymptotically far away from the hole. In the context of a neutron star, we have neutrons with at most $v_F$ velocity and $p_{\rm DB}\lesssim p_F$. Wave-like effects therefore become non-negligible for $M_{\rm BH}$ below a critical value
\begin{equation} \label{eq:unruh_applies}
    M_{\rm BH}<M_{\rm Unruh}\equiv\frac{\pi M_P^2}{p_F}=\sdmedit{2.1}\times 10^{12}\ {\rm kg}.
\end{equation}
Because these neutrons are being absorbed, they are being removed from the Fermi sea, so \sdmedit{
Pauli blocking does not inhibit their removal from the thermal bath of the neutron star}.

Note that there is some evidence that NS cores might consist of quark matter \cite{Annala:2019puf}; if that is the case, the Fermi momentum of the constituent quarks will be of the same order (up to a factor of ${\cal O}(3^{1/3})$) as that of neutrons, and the corresponding de Broglie wavelength also comparable (in fact, slightly smaller). 

We note that current limits to PBH masses from evaporation \cite{Coogan:2020tuf} constrain the PBH masses, for 100\% of the DM in PBH, to be larger than $10^{14}$ kg; the PBH mass falls in the range where Unruh's treatment is necessary only if $f_{\rm PBH}\lesssim 10^{-5}$ \cite{Coogan:2020tuf, Carr:2020gox}, in which case capture is very unlikely; however, one should treat this issue with care when discussing possible disruption of NS by accreting PBH (see e.g. \cite{Pani:2014rca}, and the ensuing debate in e.g. \cite{Montero_Camacho_2019}).


\section{Absorption rates}\label{sec:xs}
The Bondi-Hoyle absorption cross section generalizes the classical Hoyle-Lyttleton result \cite{1939PCPS...35..405H} for the accretion of \sdmedit{fluid-like flux of particles} of density $\rho$ by a star of mass $M$ moving at a steady asymptotic speed $v$,
\begin{equation}
\left(\frac{dM}{dt}\right)_{\rm HL}=\pi \zeta_{\rm HL}^2v\rho=\frac{4\pi G^2M^2\rho}{v^3},
\end{equation}
where $\zeta_{\rm HL}$ is the Hoyle-Lyttleton radius, corresponding to the maximal impact parameter yielding capture. Augmenting the Hoyle-Lyttleton treatment with fluid effects, but maintaining the assumption that the accreted particles be massless and point-like, and indicating with $c_s$ the sound speed of the fluid being accreted, gives the classic Bondi-Hoyle result \cite{1944MNRAS.104..273B,Bondi:1952ni},
\begin{equation}\label{eq:BH}
\left(\frac{dM}{dt}\right)_{\rm BH}=\frac{4\pi \lambda_s(\gamma) G^2M^2\rho}{\left(c_s^2+v^2\right)^{3/2}}.
\end{equation}
Given an equation of state $P=K\rho^\gamma$ the appropriate {\em accretion constant} $\lambda_s(\gamma)$ can be calculated using a polytropic equation of state \cite{Shapiro:1983du}; it is equal to $\lambda_s(5/3)=0.25$ in the case of degenerate matter. \sdmedit{The sound speed $c_s$ is sensitive to the equation of state, but is in the range $c_s^2 \simeq 0.2-0.4$ for the densest parts of the star \cite{Balberg:2000xu}}.

In the limit where the particles being accreted are neither massless (rather, they have mass $m$) nor point-like and possess a quantum wavelength (de Broglie wavelength) larger than the Schwarzschild radius of the accreting mass $M$, the absorption cross section was computed in Ref.~\cite{Unruh:1976fm}. It reads, for the case of a Dirac particle \sdmedit{with velocity $v$},
\begin{equation} \label{eq:unruh-rate}
    \left(\frac{dM}{dt}\right)_{\rm U}=\sigma_U(M,m,v)\rho v,
\end{equation}
with
\[
\sigma_U(M,m,v)=\frac{2\pi G^2M^2}{v}\frac{\xi}{1-e^{-\xi}}
\]
and $\xi$ defined as
\[
\xi=2\pi GMm\frac{1+v^2}{v\sqrt{1-v^2}}=\pi\frac{1+v^2}{v^2\sqrt{1-v^2}}\frac{R_{\rm Schw}}{\lambda_{\rm DB}}
\]
with $R_{\rm Schw}=2GM$ the Schwarzschild radius (in natural units) and $\lambda_{\rm DB}=1/(mv)$. Ref.~\cite{Unruh:1976fm} assumes $R_{\rm S}/\lambda_{\rm DB}\ll1$. Note that $\xi\to\infty$ as $v\to0$ and $v\to 1$.

The mass accretion rate for neutron absorption via the Unruh absorption cross section as a function of the BH mass is 
\begin{equation}
    \left(\frac{dM}{dt}\right)_{\rm U}(M)=m_n n_n \int_0^{1}dv  f_F(v)\sdmedit{v}\sigma_U(M,m_n,v),
\end{equation}
where $m_n$ is the neutron mass. Dark matter accretion is generally negligible in the growth of the black hole \cite{McDermott:2011jp} (however, it can be important in the Earth or the Sun \cite{Acevedo:2020gro}, in white dwarfs \cite{Acevedo:2019gre}, and even in the case of NS, in some corners of parameter space \cite{Bramante:2013hn}. Numerically, we find 
\begin{equation}\label{eq:approx}
    \left(\frac{dM}{dt}\right)_{\rm U}(M)\simeq \begin{cases}  10^{-38} \left(\frac{M}{\rm kg}\right)^3\frac{\rm kg}{\rm sec} & M\gtrsim 10^{10} {\rm \, kg} \\ 10^{-28} \left(\frac{M}{\rm kg}\right)^2\frac{\rm kg}{\rm sec} & M\lesssim 10^{10} {\rm \, kg} \end{cases},
\end{equation}
which reveals that the Unruh and Bondi-Hoyle rates scale similarly at low black hole mass but not at large black hole mass. \sdmedit{The transition between these regimes is at $M \simeq 10^{10}$ kg.}

\sdmedit{Let us now discuss the range of validity of, respectively, the quantum absorption Unruh picture and the Bondi-Hoyle picture. The Unruh picture assumes that (i) $R_{\rm Schw}=2GM_{\rm BH}<\lambda_{\rm dB}=1/p$ and that (ii) the infalling particle is freely falling. On the other hand, the Bondi-Hoyle picture assumes that the mean-free path of neutrons, $\lambda_{\rm mfp}\equiv 1/(n_n \sigma_{nn})$, is smaller than the Bondi radius, $R_{\rm Bondi}\equiv R_{\rm Schw}/c_s^2$. Note that the assumption that infalling particles are freely falling is effectively the reverse of this latter condition, i.e. particles are effectively ``blind'' to each other during infall if $\lambda_{\rm mfp}>R_{\rm Bondi}$.
}

{We therefore have three ranges for the black hole masses corresponding to three different accretion pictures:
\begin{enumerate}
    \item $R_{\rm Schw}>\lambda_{\rm mfp} c_s^2$ corresponds to the Bondi-Hoyle fluid-like accretion regime;
    \item $R_{\rm Schw}<\lambda_{\rm mfp} c_s^2$ {\em and} $R_{\rm Schw}<\lambda_{\rm dB}$ corresponds to the Unruh picture;
    \item if $R_{\rm Schw}<\lambda_{\rm mfp} c_s^2$ but $R_{\rm Schw}\gtrsim\lambda_{\rm dB}$ we are in an intermediate regime where the Bondi picture fails, but where the absorption cross section as approximated by the Unruh result is also inapplicable. In this case,  the correct picture is that of freely-falling particles being {\em classically} absorbed by the hole;  the correct cross section is therefore the classical absorption cross section \cite{1962upei.book.....B}
{\small
\begin{equation}
    \sigma_C(M,v)=\frac{\pi G^2M^2}{v^2}\left(\frac{\left(8(1-v^2)\right)^3}{4(1-4v^2+\sqrt{1+8v^2})(3-\sqrt{1+8v^2})^2}\right).
\end{equation}} 
\end{enumerate}

We now estimate the masses corresponding to the three regimes listed above. First, since we are concerned with the very central region of the neutron star, we shall assume a relatively large sound speed, $c^2_s = 0.3$. Second, we use an effective nucleon mass of $m_N^*=750$ MeV to account for the effects of nuclear forces in this high-density environment \cite{Rrapaj:2015zba, Drischler:2016cpy, Li:2018lpy}. Lastly we assume $\sigma_{nn} = 10\ {\rm mb}$, which is a representative value at the momentum $k=p_F$ for the effective range parameters $a_{nn} = -18.5{\rm \,fm}$ and $r_{nn}=2.75{\rm \,fm}$ \cite{Gardestig:2009ya, Gobel:2021pvw}. With these values, we obtain $\lambda_{\rm mfp}c^2_s \simeq 1$ fm, and $\lambda_{\rm dB}\simeq (m_N^* \langle v \rangle_F )^{-1} \simeq 0.88$ fm\footnote{\steedit{We note that it is plausible that our estimate of the range of applicability of the fluid behavior might actually be overly conservative: a criterion based on neutron relaxation times \cite{1976ApJ...206..218F}, involving neutrons further from the Fermi surface, might effectively yield a larger mean free path \cite{Shternin:2010qi}, implying, as a result, an even broader range of validity for the effects under discussion here. Additional effects associated with superfluid behavior in the neutron star core \cite{Alford:2007xm} might additionally affect our conclusions.}}. The corresponding mass ranges are thus:
\begin{enumerate}
    \item the black hole is in the Bondi-Hoyle regime if $R_{\rm Schw}>\lambda_{\rm mfp} c_s^2$, which is satisfied if $M>M_{\rm Bondi} \equiv 6.7\times 10^{11}$ kg;
    \item since $\lambda_{\rm mfp} c_s^2>\lambda_{\rm dB}$, the black hole is in the Unruh regime if $M<M_{\rm Unruh} \equiv 5.9\times 10^{11}$ kg; and
    \item the black hole is in the classical regime in the narrow range of masses for which $M_{\rm Unruh}< M< M_{\rm Bondi}$.
\end{enumerate}
}


\section{Neutron Star Lifetime}\label{sec:lifetimes}
In addition to accretion, the black hole mass changes because of Hawking evaporation, at a rate given by \cite{MacGibbon:1991tj} 
\begin{equation}\label{eq:evaporation}
\left(\frac{dM}{dt}\right)_{\rm H}(M)\simeq-5\times 10^{16}f(M)\left(\frac{\rm kg }{M}\right)^2\ \frac{\rm kg}{\rm s},
\end{equation}
where $f(M)$ is a function of the degrees of freedom kinematically available for evaporation: only those particles for which the Hawking temperature $T_{\rm H}\gtrsim m$, where $m$ is the particle that the black hole evaporates into, can be produced by the black hole. For $M\sim 10^{9}$ kg, $T_{\rm H}\sim 10$ GeV and $f(M)\simeq 15$, while for $M\sim 10^{13}$ kg, $T_{\rm H}\sim 1$ MeV and $f(M)\simeq 2$. We use the full form for $f(M)$ as given in \cite{MacGibbon:1991tj, Carr:2020gox}.

The black hole mass as a function of time is given in general by
\begin{equation} \label{eq:diffeq-M}
    M(t) = \int_{t_0}^t dt \left[\left(\frac{dM}{dt}\right)_{\rm acc} + \left(\frac{dM}{dt}\right)_{\rm H} \right],
\end{equation}
where $\left(dM/dt\right)_{\rm acc}$ connotes the appropriate accretion rate. Clearly, because of the different signs of the rates in Eq.~\eqref{eq:diffeq-M}, there is a critical rate below which the black hole mass inexorably falls.
We find that black hole evaporation dominates over matter accretion for hole masses less than 
\begin{equation} \label{eq:Mcrit}
    M_{\rm crit}\simeq 1.6\times 10^{11}\ {\rm kg};
\end{equation}
for smaller masses, the black hole evaporates rather quickly. For instance, for an initial mass of $10^{11}$ kg, thus barely below $M_{\rm crit}$, the hole evaporates in $4\times 10^{13}$ sec, which is only a thousandth the age of observed nearby NS such as PSR J0437-4715 and PSR J2124-3358, both on the order of $10^{10}$ years \cite{Manchester:2004bp}. \steedit{We note that most pulsars are younger than those mentioned above, with spin-down ages closer to $10^6-10^7$ years \cite{Manchester:2004bp}; however, since here we are interested in the steady accretion of dark matter on the neutron star, subsequently triggering the formation of a black hole, longer-lived systems are those of interest to us.}
We find that a good approximate fit for small masses is
\[
\tau_{\rm evap}(M)\simeq 8\times 10^9\ {\rm sec}\ \left(\frac{M}{10^{10}\ {\rm kg}}\right)^3,\quad (M<M_{\rm crit})
.
\]Note that unlike the case of evaporation of a black hole inside the Earth or the Sun \cite{Acevedo:2020gro}, evaporation inside a NS is not expected to yield any observable signature: comparing the rest-mass energy of the largest hole that would evaporate quicker than accrete, $M\sim M_{\rm crit}\simeq 8\times 10^{34}$ ergs,  with the {\em lower limit} to the specific heat of a NS, $c_{NS}\gtrsim 2\times 10^{36} {\rm ergs}/$K \cite{Cumming:2016weq} makes it clear that the deposited heat would never yield a detectable temperature change to the NS. Nevertheless, it is possible that this sudden deposition of energy in the NS core will have a transient effect such as a glitch. We also estimate that the neutrino mean free path inside a NS is too short for neutrinos to escape
$$
\lambda_\nu\simeq \frac{1}{n_n \sigma_{n\nu}}\simeq \frac{1}{n_n G_F^2 E_\nu^2}\simeq 2\times 10^{-8}\ {\rm cm}\left(\frac{\rm GeV}{E}\right)^2
$$
so that the predicted flux would be too small to be detectable above the atmospheric neutrino background ({\it e.g.,} \cite{Acevedo:2020gro}, Fig.~6).

For initial black hole masses larger than $M_{\rm crit}$, we can determine the neutron star lifetime via
\begin{equation} \label{eq:diffeq-t}
    \tau(M_0) = \int_{M_0}^{M_{\rm NS}} \frac{dM}{\left(\frac{dM}{dt}\right)_{\rm acc} + \left(\frac{dM}{dt}\right)_{\rm H}},
\end{equation}
where $M_0$ is the initial BH mass and $M_{\rm NS} \simeq 1.5 {\rm M}_\odot$ is the neutron star mass. The resulting NS lifetime $\tau(M)$, using the full numerical solution, is
{
\begin{equation}
    \frac{\tau(M)}{{\rm Myr}}\simeq \begin{cases}  0.6\left(\frac{M_{\rm Unruh}}{M}\right)^2, & M_{\rm crit}<M<M_{\rm Unruh} 
    \\ 100\left(\frac{M_{\rm Unruh}}{M}\right), & M>M_{\rm Unruh}. \end{cases}
\end{equation}}
Because $M_{\rm crit}$ given in Eq.~\eqref{eq:Mcrit} is of order $10^{11}$ kg, the neutron star destruction rate is shorter than $\tau_{\rm NS}$ if the black hole mass is sufficiently large to avoid evaporating.

\begin{figure}
    \centering
    \includegraphics[width=0.49\textwidth]{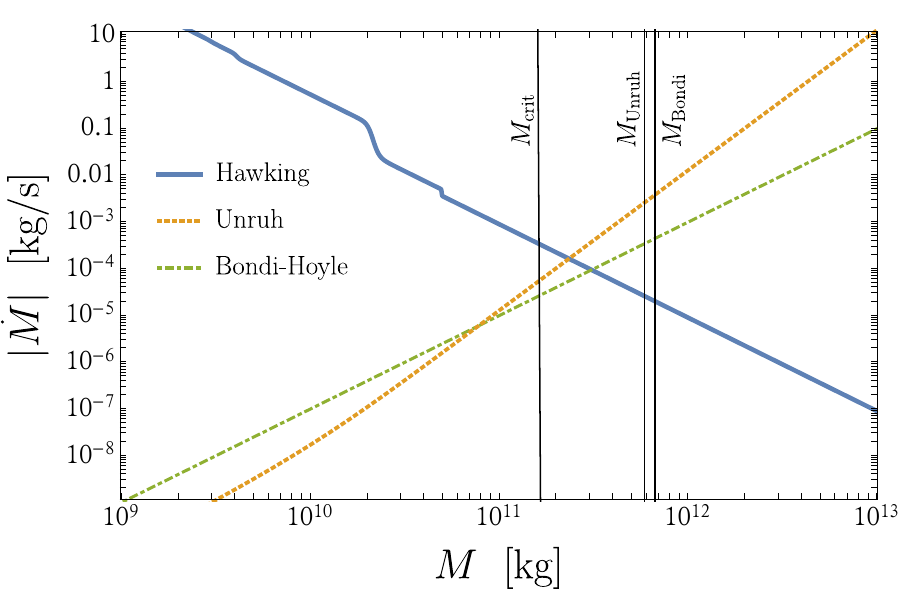}
    \caption{\textbf{Accretion and evaporation rates.} We show the accretion and evaporation rates, in units of kg/sec, for the evaporation rate (Eq.~(\ref{eq:evaporation}), blue line); the Unruh quantum accretion process (orange dotted line)
    ; and the Bondi-Hoyle accretion rate (Eq.~(\ref{eq:BH}), green dot-dashed line \sdmedit{(lying below the Unruh accretion line at large masses)}). \sdmedit{The black hole evaporates faster than it grows if $M<M_{\rm crit}$; it accretes mass according to the Unruh rate for $M < M_{\rm Unruh}$ and according to the Bondi-Hoyle rate for $M>M_{\rm Bondi}$. See text for details.}}
    \label{fig:rates}
\end{figure}

\section{Black Holes from Dark Matter Collapse in Neutron Stars}\label{sec:bhcollapse}
The mass of the black hole formed from dark matter collapse is the maximum between the largest mass supported by quantum pressure and the largest self gravitating mass, \cite{Acevedo:2020gro}
\begin{equation}\label{eq:sg}
M_{\rm sg}=\sqrt{\frac{3T^3}{\pi G_N^3 m^3\rho}}\simeq 134\ {\rm kg}\ \left(\frac{T}{10^5\ {\rm K}}\right)^{3/2}\left(\frac{\rm GeV}{m}\right)^{3/2}.
\end{equation}
The critical particle number $N$ that leads to exceeding quantum pressure support against gravitational collapse depends on the spin of the dark matter.

In the case of fermions, the onset of the gravitational collapse occurs when the potential energy of the dark matter exceeds the Fermi energy, and therefore Pauli blocking cannot prevent the collapse anymore:
\begin{equation}
    \frac{GN_{\rm max}m_f^2}{r}=E_F=\left(\frac{3\pi^2N}{V}\right)^{1/3}=\left(\frac{9\pi}{4}\right)^{1/3}\frac{N_{\rm max}^{1/3}}{r}.
\end{equation}
The radius of the self-gravitating sphere drops out of this expression, and thus the BH mass is
\begin{equation} \label{eq:fermion-MBH}
    M_{\rm BH}^f=N_{\rm max}^f m_f \simeq 9\times 10^{30}\ {\rm kg}\left(\frac{\rm GeV}{m}\right)^2.
\end{equation}
This expression holds for self-gravitating, non-interacting fermions. Corrections due to self-interactions are important for the case of neutrons, and the maximum neutron star mass is not precisely known for this reason \cite{Shapiro:1983du}.

In the case of bosons, the energy for a single particle is
\begin{equation}
    E\sim-\frac{GNm_b^2}{R}+\frac{1}{2m_bR^2}-\frac{\lambda N}{32\pi m_b^2 R^3},
\end{equation}
where the second term stems from the particle kinetic energy due to the uncertainty principle and the final term is due to the particle self-interactions. As we discuss in more detail in the Appendix, the maximum number of bosons that are stable against gravitational collapse are
\begin{equation}
    N_{\rm max}^b = \left(\frac{M_{\rm Pl}}{m_b}\right)^2 \sqrt{ \frac{17}{20} \left(1 - \frac{3\lambda M_{\rm Pl}^2}{34 \pi m_b^2} \right)}.
\end{equation}
The black hole mass that is obtained if the number of particles exceeds this value is
\begin{equation} \label{eq:bos-Mmax}
    M_{\rm max}^b\simeq 2.5\times 10^{14}\ {\rm kg}\frac{\rm GeV}{m_b} \sqrt{1-4\times10^{36} \lambda \left(\frac{\rm GeV}{m_b}\right)^2}.
\end{equation}
Clearly, the sign and the magnitude of $\lambda$ matter very much for the mass of the black hole. If $\lambda$ is positive, corresponding to an attractive self-interaction, only extremely small values of this coupling are possible in a stable system. Here, we will focus on a few representative cases: $\lambda=0$, which is possible if $\phi$ is exactly protected by a large symmetry group; $\lambda= -(m/f)^2,$ which is the first term in the expansion of some non-analytic potentials motivated by quantum gravity \cite{Choi:2019mva}, where $f$ is a ``decay constant'' corresponding to massive modes for which we take $f=10^{10}$ GeV and $10^{12}$ GeV; and constant values $\lambda=-10^{-2}$ and $\lambda= - 0.1^2 (1/16\pi^2)^2 \simeq -\times10^{-7}$, which is of the correct size for a loop-induced self coupling arising from integrating out a perturbatively coupled scalar \cite{Bell:2013xk}. Other realistic models with small repulsive couplings were obtained in \cite{Fan:2016rda, Croon:2018ybs}.

Finally, we note that accumulated bosonic dark matter can form a Bose-Einstein condensate (BEC) \cite{McDermott:2011jp}. This can trigger black hole formation from the condensate subcomponent of the dark matter rather than the entire thermal population. The fraction of dark matter particles in the BEC if the star is below the critical temperature is formally $N_{\rm BEC}/N^b = \Theta(T_{\rm crit} - T_c) [1-(T_c/T_{\rm crit})^{3/2}]$, where $T_c$ is the core temperature of the star. We emphasize here that the dependence on temperature is dominated by the step function: if the temperature in the core of the star is below $T_{\rm crit}$, the majority of the particles are in the BEC, unless the temperature is extremely close to the phase transition. Thus, we approximate the mass of the BEC as zero if $T_c > T_{\rm crit}$ and as $m_X N^b$ if $T_c < T_{\rm crit}$. The critical temperature of a non-interacting bosonic system in a square-well potential is\footnote{The presence of self interactions and the harmonic (rather than square-well) nature of the potential after onset of self-gravitation can in fact increase the critical temperature and thus make the condensation of the ground state moderately more favorable \cite{Huang:1999zz, Jamison:2013yya}, but these changes are at or below the order of magnitude level, and we omit them here for simplicity.} $T_{\rm crit} = \frac{2\pi}m \left[ \frac {3 N^b }{4\pi \zeta(3/2) r_{\rm th}^3} \right]^{2/3}$. The radius inside of which the thermalized DM particles are distributed scales like $r_{\rm th} \propto \sqrt{T_c/m_X}$ and the total number of particles scales like $N^b \propto \rho_X \sigma_{XN} t [\max({\rm GeV},m_X)]^{-1}$, where $\rho_X$ is the DM in the vicinity of the NS, $\sigma_{XN}$ is the nucleon-$X$ scattering cross section, and $t$ is the age of the star \cite{McDermott:2011jp}. For convenience, we define a scaling function $g_{\rm BEC} = \frac{\rho_X}{\rm GeV/cm^3} \frac{\sigma_{XN}}{10^{-45} {\rm cm}^2} \frac{t}{\rm 10Gyr} e_{\rm cap}$ which accounts for the age of the neutron star, the conditions of the DM in its vicinity, and the efficiency of capture $ e_{\rm cap}$, which can be small when $m_{\rm DM}$ becomes too large \cite{McDermott:2011jp, Bramante:2013hn, Acevedo:2020gro}. Combining all of these ingredients and plugging in numbers from \cite{McDermott:2011jp}, we find that the mass of the BEC is 
\begin{eqnarray} \label{eq:mbec}
    &&M_{\rm BEC}^b \simeq 2\times10^{16}\min\left( \tfrac{m}{\rm GeV}, 1 \right) g_{\rm BEC} \,{\rm kg} 
    \\ &&~~{\rm if }\, \max\left( \tfrac{m}{\rm GeV},1 \right) < \left( \tfrac{13}{T_6} \right)^3 \left( \tfrac{n_N}{0.3 {\rm fm}^{-3}} \right)^{3/2} g_{\rm BEC} , \nonumber
\end{eqnarray}
where we have defined $T_6 = T/10^6{\rm \,K}$.
Because we are primarily interested in this work with the behavior of the black hole, rather than the constraints on the dark matter parameter space, we will set $g_{\rm BEC}=1$, since this will be true after a sufficiently long time regardless of the environment.

Since neutron star temperatures fall to around $10^6$ K after approximately a Myr and stay stable at that order of magnitude for roughly a Gyr \cite{Hamaguchi:2019oev}, Eq.~\eqref{eq:mbec} indicates that a reasonable expectation is that BEC formation will be important for bosonic dark matter masses of order a TeV. However, the entire thermal distribution of captured particles may exceed $M_{\rm max}^b$ given in Eq.~\eqref{eq:bos-Mmax} before condensation is triggered. Thus, the black hole mass that we expect from accumulation of bosonic particles is
\begin{equation} \label{eq:bos-MBH}
    M_{\rm BH}^b = \min(M_{\rm BEC}^b, M_{\rm max}^b).
\end{equation}
This is a function of time through the dependence of Eq.~\eqref{eq:mbec} on $g_{\rm BEC}$, which we are setting to 1 for illustrative purposes.

\begin{figure}[t]
    \centering
    \includegraphics[width=0.49\textwidth]{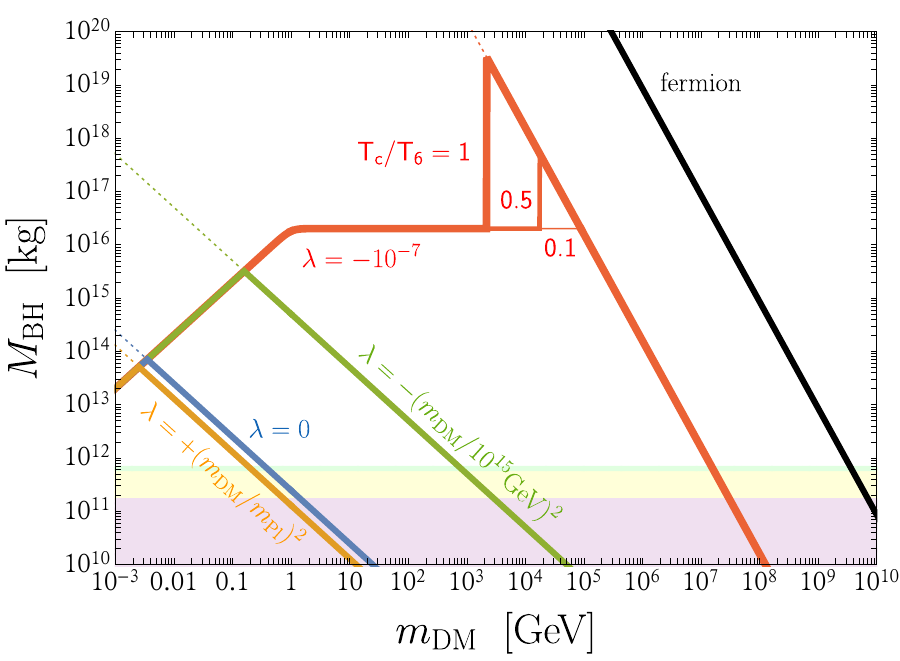}
    \caption{\textbf{Black hole masses for various dark matter models.} We show the predicted mass of black holes formed by accreting a critical number of dark matter particles inside a NS. The fermion line (black) from Eq.~\eqref{eq:fermion-MBH} is model-independent. The boson lines from Eq.~\eqref{eq:bos-MBH} show the sensitivity to the self-interaction coupling and the presence of a BEC. For a constant repulsive self interaction, which we illustrate with $\lambda=-10^{-7}$ (red), the black hole mass at large $m_{\rm DM}$ is similar to the fermion case, but smaller by a factor $\sqrt{-\lambda}$. At lower masses, the slope of the $\lambda=-10^{-7}$ line differs from that of the fermion line by virtue of BEC formation, where we have assumed $T_c = 1 (0.5) [0.1] \times 10^6$ K for the thick (medium) [thin] weight lines. The lines $\lambda=0$ (blue), $\lambda = -(m_{\rm DM}/10^{15} {\rm GeV})^2$ (green), and $\lambda = +(m_{\rm DM}/m_{\rm Pl})^2$ (orange), where we define the reduced Planck constant $m_{\rm Pl} = M_{\rm Pl}/\sqrt{8\pi}$, are parallel to one another. The dotted lines extending above each solid bosonic line show the black hole mass if BEC formation is neglected. 
    }
    \label{fig:moneyplot}
\end{figure}

Figure \ref{fig:moneyplot} shows the mass of the BH as a function of the dark matter mass. The fermion line, given by Eq.~\eqref{eq:fermion-MBH}, is appropriate given the minimal assumptions that the dark matter is able to self-gravitate and is not strongly self-interacting. The boson lines are more sensitive to the model parameters. We attempt to demonstrate the sensitivity of the final black hole mass on the self-interaction coupling and the presence of a BEC. When bosonic dark matter has a constant repulsive self interaction, which we illustrate with $\lambda=-10^{-7}$, the black hole mass at large $m_{\rm DM}$ is similar to the fermion case, though smaller by a factor $\sqrt{-\lambda}$. At lower masses, even this case diverges from the fermion line by virtue of BEC formation, however. The value of $m_{\rm DM}$ at which BEC formation becomes important depends on the core temperature of the NS. For illustration purposes, we assume $T_c = 10 (5) [1] \times 10^6$ K as representative values, with divergences due to the BEC phase transition from roughly 2 TeV, to 20 TeV, to no divergence, respectively. The other noticeable feature in the red line occurs at $m_{\rm DM}=1$ GeV, because the efficiency of capture of lower-mass dark matter particles falls due to Pauli blocking. Finally, we show lines $\lambda=0$ (blue), $\lambda = -(m_{\rm DM}/10^{15} {\rm GeV})^2$ (green), and $\lambda = +(m_{\rm DM}/m_{\rm Pl})^2$ (orange), where we define the reduced Planck constant $m_{\rm Pl} = M_{\rm Pl}/\sqrt{8\pi}$. These are motivated by axion models; they are parallel. We note that there are no self-gravitating solutions at all for $\lambda > + 40\pi m_{\rm DM}^2/3M_{\rm Pl}^2.$ The dotted lines extending above each solid bosonic line show the black hole mass if BEC formation is neglected.

\section{Summary \& conclusions}\label{sec:summary}
When the quantum size of neutrons exceeds the Schwarzschild radius of a black hole at the center of a neutron star, accretion cannot be described with the Bondi-Hoyle picture; rather, it should be described by an appropriate cross section that accounts for both the space-time geometry of the black hole, and the quantum nature of the particles being accreted.

Here, we corrected the predictions for neutron star destruction by black holes formed by non-annihilating dark matter accumulating at the neutron star interior using the correct capture cross section for light black holes. \steedit{While the key results in the existing literature are not quantitatively dramatically affected, we find a significant change in the minimal critical seed black hole mass  necessary to prevent black hole evaporation and to trigger the disruption of neutron stars, and in the resulting predicted neutron star lifetime}.

Future work will tackle the complex problem of fermion accretion onto Schwarzschild black holes \steedit{(or onto spinning black holes more generally)} at finite temperature \steedit{as well as possibly applying the density matrix formalism to treat quantum accretion properly. Finally, we also mention that for certain dissipative dark matter models, dark matter accretion can also play a role in determining the black hole growth rate.} \cite{inprep}.

\vspace*{.5cm}

\begin{acknowledgments}
We gratefully acknowledge conversations with Joseph Bramante, Joachim Kopp, and Haibo Yu.
SP is partly supported by the U.S.\ Department of Energy grant number de-sc0010107.
SDM appreciates conversations with Djuna Croon and would like to thank the GGI for hospitality and Ken van Tilburg for pertinent discussions there. SDM is supported by the Fermi Research Alliance, LLC under Contract No.~De-AC02-07CH11359 with the United States Department of Energy, Office of High Energy Physics.
\end{acknowledgments}

\bibliographystyle{apsrev4-1}
\bibliography{references}

\end{document}